\documentclass[prd,twocolumn,superscriptaddress,nofootinbib,noshowpacs,preprintnumbers,longbibliography,floatfix]{revtex4-2}

\usepackage{graphicx}
\usepackage{dcolumn}
\usepackage{bm}
\usepackage{amsmath}
\usepackage{amsfonts}
\usepackage{amssymb}
\usepackage{dsfont}
\usepackage{color}
\usepackage{xcolor}
\usepackage{siunitx}
\usepackage{slashed}
\usepackage[colorlinks=true,
    linkcolor=blue,
    filecolor=blue,      
    urlcolor=blue,
    citecolor = blue]{hyperref}
\usepackage{braket}
\usepackage{mathtools}
\usepackage[mathlines]{lineno}
\usepackage[normalem]{ulem}
\usepackage{scalerel}
\usepackage{ bbold }
\usepackage{inputenc}
\usepackage[T1]{fontenc}

\graphicspath{{./figures/}}

\definecolor{mmagenta}{cmyk}{0,1,0,0.12}

\begin{document}

\title{Nonperturbative signatures of fractons in the twisted multiflavor Schwinger Model}

\date{\today}

\author{Pavel P. Popov}
\email{pavel.popov@icfo.eu}
\affiliation{ICFO - Institut de Ciencies Fotoniques, The Barcelona Institute of Science and Technology, Av. Carl Friedrich Gauss 3, 08860 Castelldefels (Barcelona), Spain}

\author{Valentin Kasper}
\affiliation{ICFO - Institut de Ciencies Fotoniques, The Barcelona Institute of Science and Technology, Av. Carl Friedrich Gauss 3, 08860 Castelldefels (Barcelona), Spain}

\author{Maciej Lewenstein}
\affiliation{ICFO - Institut de Ciencies Fotoniques, The Barcelona Institute of Science and Technology, Av. Carl Friedrich Gauss 3, 08860 Castelldefels (Barcelona), Spain}
\affiliation{ICREA, Pg. Lluis Companys 23, 08010 Barcelona, Spain}

\author{Erez Zohar}
\affiliation{Racah Institute of Physics, The Hebrew University of Jerusalem, Givat Ram, Jerusalem 91904, Israel}

\author{Paolo Stornati}
\affiliation{ICFO - Institut de Ciencies Fotoniques, The Barcelona Institute of Science and Technology, Av. Carl Friedrich Gauss 3, 08860 Castelldefels (Barcelona), Spain}

\author{Philipp Hauke}
\affiliation{Pitaevskii BEC Center and Department of Physics, University  of  Trento,  Via Sommarive 14, I-38123 Trento, Italy}
\affiliation{INFN-TIFPA, Trento Institute for Fundamental Physics and Applications, Trento, Italy}

\begin{abstract}

Gauge-field configurations with nontrivial topology have profound consequences for the physics of Abelian and non-Abelian gauge theories. Over time, arguments have been gathering for the existence of gauge-field configurations with fractional topological charge, called fractons. Ground-state properties of gauge theories can drastically change in presence of fractons in the path integral. However, understanding the origin of such fractons is usually restricted to semiclassical argumentation. 
Here, we show that fractons persist in strongly correlated many-body systems, using the multiflavor Schwinger model of quantum electrodynamics as a paradigm example. 
Through detailed numerical tensor-network analysis, we find strong fracton signatures even in highly discretized lattice models, at sizes that are implementable on already existing quantum-simulation devices. 
Our work sheds light on how the nontrivial topology of gauge theories persists in challenging nonperturbative regimes, and it shows a path forward to probing it in tabletop experiments.

\end{abstract}

\maketitle

\section{Introduction}The topology of the vacuum plays a crucial role in some of the most fundamental theories of nature, including gauge field theories \cite{PhysRev122345} and supersymmetric models \cite{Amati_1988ft}.
It governs the subtle mechanisms behind such phenomena as charge confinement and chiral symmetry breaking, and topological theta vacua \cite{Belavin1975} are intrinsically connected to the strong CP problem in quantum chromodynamics (QCD) in 3+1D \cite{PhysRevLett_67_2765}. 
Vacuum sectors with distinct topological charge arise naturally from gauge-field configurations with different windings~\cite{Callan1976}. 
While naive analysis may suggest the relevance of only sectors with integer windings, configurations with deconfined \emph{fractional topological charge}---called \emph{fractons}---can exist \cite{tHooft_1981,Shifman1994} and are in fact vital in resolving paradoxes related to nonvanishing gluino condensates in supersymmetric Yang--Mills theory \cite{Cohen_1984,davies1999gluino}.
Similarly, in non-Abelian gauge theories they can explain the mechanisms behind the formation of a fermion condensate where considerations based on only integer topological charges fall short \cite{Leutwyler1992,Shifman1994}. 
However, existing insights into the importance of fractons derive from exactly solvable models or semiclassical arguments. It is not clear if and how signatures of fractons persist in nonperturbative regimes of strongly correlated theories.

In this work, we demonstrate the presence of fractional gauge-field configurations in the full nonperturbative quantum many-body regime of a paradigmatic gauge theory. 
Our study dives into the Schwinger model of quantum electrodynamics (QED) with two fermionic flavors, which captures the essence of more complicated quantum field theories, such as QCD, in a simpler, more tractable form. 
The Schwinger model is a well-established paradigm to facilitate insights into phenomena such as charge confinement, topological theta vacua, and chiral symmetry breaking \cite{Melnikov2000}.
A more recent condensed-matter perspective has also revealed new phenomena \cite{funcke2020topological,PhysRevA_90_042305,magnifico2020real} such as quantum many body scars \cite{PhysRevLett_124_180602,Banerjee2021_scars,Halimeh2023,Sau2024,Desaules2023}, disorder-free localization \cite{Brennes2018,PhysRevB.106.174305}, or dynamical topological phase transitions \cite{PhysRevLett.122.050403,PRXQuantum.4.030323,e25040608}. 
The multiflavor version permits for flavor-twisted boundary conditions, which fundamentally modify the symmetry properties of the theory. 
Analytical calculations at 
perturbatively small fermion mass \cite{Shifman1994,Misumi2019} have shown that
these lead to the deconfinement of fractons. They become visible through a nonzero chiral condensate as well as a fractional dependence of the ground state on the topological theta angle. 

Based on tensor network (TN) calculations and exact diagonalization, we demonstrate the persistence of these fracton signatures in the nonperturbative  regimes of significant rest mass as well as sizable lattice spacing. 
In the limit of vanishing rest mass, even coarse lattice discretizations  quantitatively recover the perturbative continuum predictions. 
Our numerics demonstrates the robustness of fractons against strong quantum fluctuations, lattice artifacts, as well as a cutoff on the gauge-field Hilbert space. 
Leveraging this considerable robustness toward discretization, we propose a variational quantum algorithm for exploring fractons in a qudit quantum simulator, as has recently been demonstrated in trapped ions \cite{Ringbauer2022,Meth2023}. 
Strong fracton signatures can be observed in such a device with already existing resources, and in the future, it may enable to proceed into regimes beyond the capacities of classical numerics. 
Our results thus demonstrate the importance of fractional gauge-field configurations in the nonperturbative regime of strongly coupled models, and they present a clear avenue for probing them in quantum hardware.

In the following, we give a brief review of how fractional gauge fields arise in the continuum in the presence of flavor-twisted boundary conditions. 
The main part of this work consists of our numerical calculations, which reveal fractons in nonperturbative regimes of the truncated lattice Schwinger model. 
After an analysis of discretization effects, we discuss a variational scheme to probe fracton physics on an existing quantum-simulator platform, before presenting our conclusions.

\section{Fractons in the multiflavor Schwinger model}We consider (1+1)-dimensional QED \cite{PhysRev_128_2425,PhysRev_82_914} with $N$ fermionic flavors, living on a cylinder $\mathbb{R}\times \mathbb{S}_L$ that is closed in the spatial direction. We denote its circumference as $L$ and its volume as $V = \mathbb{R}\times L$. The continuum action of the theory is (for details, see Appendix~\ref{app:path_integral})
\begin{align}
    S = \int_{V}d^2x \bigg\{&-\frac{1}{4}F_{\mu\nu}^2 +i\sum^N_{p=1}\bar{\psi}_p\slashed{D}\psi_p - \sum^N_{p=1}m_p\bar{\psi}_p \psi_p\bigg\}\,,
\label{eq:action_Schwinger}
\end{align}
where $D_{\mu} = \partial_{\mu}-ieA_{\mu}$ is the covariant derivative, capturing the gauge-field--matter interactions. 

Fractons are gauge-field configurations in the Euclidean path integral for which the topological invariant, the two-dimensional equivalent of the Pontryagin class,
\begin{align}
    \nu_2 = \frac{e}{4\pi}\int_V d^2x\epsilon_{\mu\nu}F_{\mu\nu}\,,
\end{align}
is a fraction. Often, the symmetry group restricts the possible values of $\nu_2$ to integers, as happens, e.g., for Abelian anyons in condensed-matter systems \cite{RevModPhys_80_1083}. For the multiflavor Schwinger model, 
deconfined fractons are allowed by the gauge group, and thus, the model provides an intriguing test bed for probing their properties.

However, even if fractons exist in a given theory, they can be difficult to observe. 
In the multiflavor Schwinger model, fractons can be revealed by imposing flavor-twisted boundary conditions for the fermionic fields on the spatial circle, $\psi_p(x+L) = e^{i\alpha_p}\psi_p(x)$, where $\alpha_p = 2\pi p/N$. For this choice of boundary conditions, the fractons become visible in the Euclidean path integral~\cite{Shifman1994}, thanks to a symmetry that emerges from the explicit violation of flavor symmetry and ``large gauge transformation''; see Appendix~\ref{app:path_integral}. One consequence of the presence of fractons is a change of periodicity of the ground state with respect to the $\theta$ angle. Conversely, such an increased periodicity can be used to detect the fractons.

Fractional gauge field configurations also influence specific observables. 
For example, the chiral condensate at zero temperature and vanishing fermion mass $m$ can be calculated through the path-integral partition function $Z(m)$ as 
\begin{align}
    \label{eq:chiralcond_fromZ}
    \langle\bar{\psi}\psi\rangle = -\frac{\partial}{\partial m}\ln Z(m)\vert{_{m = 0}}\,, \quad Z(m)\propto m^{|\nu_2|N}\,.
\end{align}
From Eq.~\eqref{eq:chiralcond_fromZ}, it follows that a nonzero chiral condensate can only be obtained due to gauge-field configurations with a fractional topological charge $\nu_2 = 1/N$. Thus, the spontaneous formation of a chiral condensate 
is another fingerprint of the existence of fractons.


By including fractons in the path integral, in the limit of vanishing rest mass one obtains analytic predictions for the full $L$-dependent chiral condensate~\cite{Shifman1994}: 
\begin{align}
\langle\bar{\psi}\psi\rangle = \sqrt{\frac{\mu e^{\gamma}}{16\pi L}}e^{-I(L,\mu)/2},
\end{align}
with $\mu^2 = Ne^2/\pi$ the photon mass, $\gamma$ Euler's constant, and $I(L,\mu)$ a Bessel function given in Appendix~\ref{app:detecting}.

The above analysis holds strictly speaking for the continuum Schwinger model in the exactly solvable limit $m/e\ll 1$. 
It is not \textit{a priori} clear that a nonperturbative regime will exhibit 
the same physics. 
For example, in the opposite limit of $e/m \ll 1$, the gauge and matter sectors decouple. 
Chiral condensation then occurs not due to fractons but due to an explicit breaking of chiral symmetry. 
This motivates the question of how the role of fractons persists into the nonperturbative regime of intermediate couplings $e/m$. 
Similarly, it remains open whether signatures of fractons carry over to strongly interacting many-body lattice versions of the Schwinger model. 
As an additional complication of such lattice models, they typically employ a finite cutoff on the gauge-field Hilbert space. This cutoff makes it impossible to define a vector potential (analogous to the difficulties in defining a phase operator) \cite{Chandrasekharan1997} and thus obstructs arguments relying on windings of gauge fields. 

In this article, we dedicate our effort on answering this question: Can we find signatures for fracton contribution to the ground state physics of a nonperturbative many-body system and, in particular, one that can be implemented in present days quantum hardware?

\begin{figure}[t!]
    \centering
    \includegraphics[width = 8.6cm]{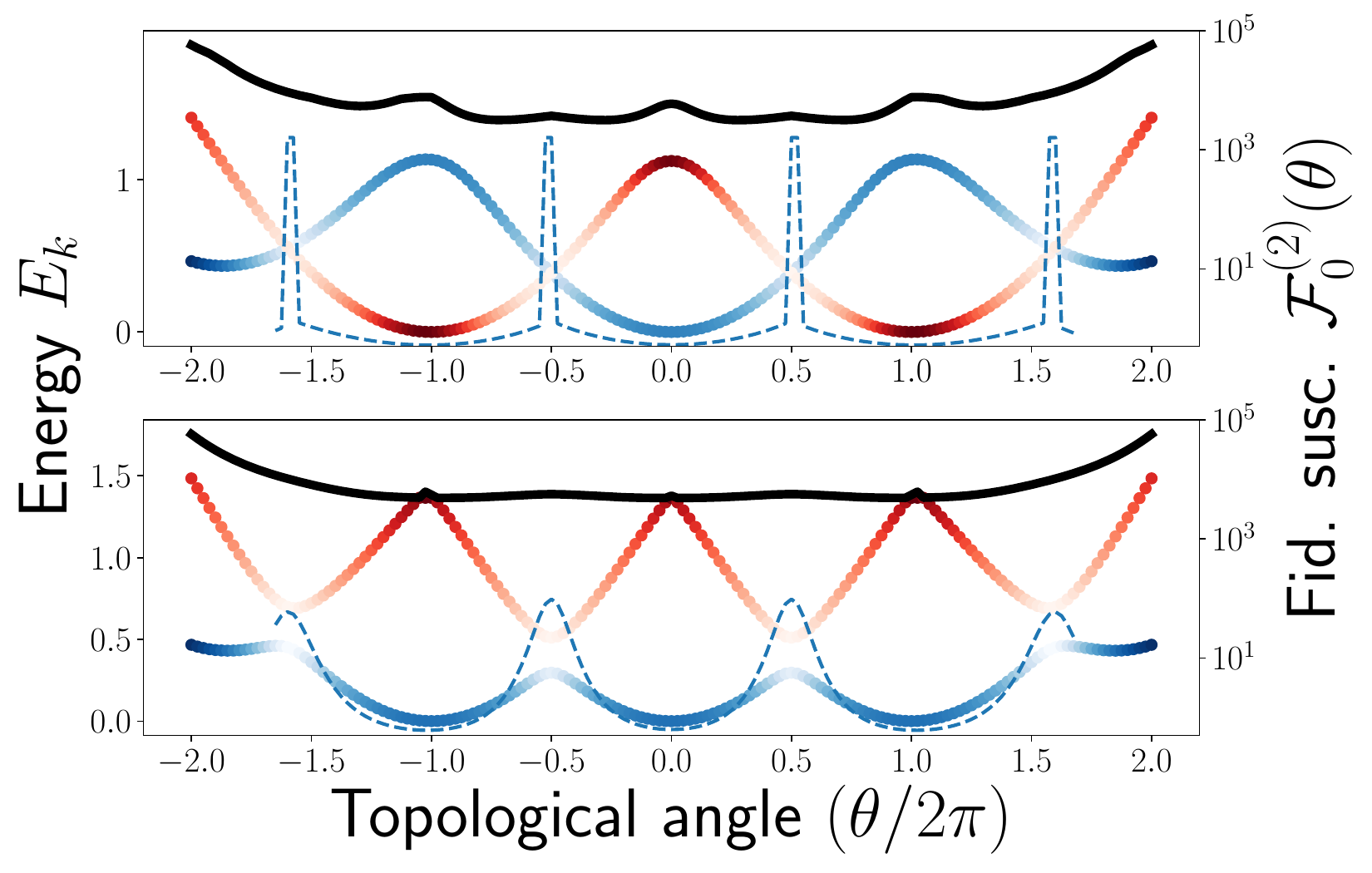}
    \caption{\textbf{Integer vs.~fractional $\theta$ dependence in the multiflavor TSM.} Lowest three energy levels of the TSM Hamiltonian (ground and first excited state:  blue and red; second excited state: black) for flavor-twisted (upper panel) and flavor-independent (lower panel) boundary conditions, as a function of the $\theta$ angle. In the former case, at $\theta = \pm \pi$, the two lowest lying states cross, and the fidelity susceptibility (blue dotted line) exhibits a strong peak. This indicates a rapid change in the properties of the ground state, suggesting that the periodicity of the ground state is $4\pi$. In contrast, in case of flavor-independent boundary conditions, avoided crossings occur at $\theta = \pm \pi$ with only a broad feature in the fidelity susceptibility, rendering the period of each energy level $2\pi$.}
    \label{fig:fractons}
\end{figure}
\section{Truncated lattice Schwinger model}

To make the continuum theory suitable for simulation techniques developed for many-body systems, we formulate it as a lattice Hamiltonian, following the prescription due to Kogut and Susskind~\cite{Kogut1975}. In the continuum, after performing a Legendre transformation on the Lagrangian in the action of Eq.~\eqref{eq:action_Schwinger} (see also Appendix~\ref{app:path_integral}), one obtains the Hamiltonian of the multiflavor Schwinger model as 
\begin{align}
    H = \sum_{p = 1}^{N}\int dx \Big[&\psi^{\dagger}_{p}(x)\gamma_0i\gamma_1\big(\partial_x+ieA_1(x)\big)\psi_{p}(x)\notag\\&+m_{p}\psi^{\dagger}_{p}(x)\psi_{p}(x)+\frac{1}{2}\big(E(x)+\theta\big)^2\Big].
\end{align}
The gauge symmetry of the Lagrangian survives in the Hamiltonian formulation as an operator constraint; the so-called Gauss's law
\begin{align}
    G(x) = \nabla E(x) - e\sum_{p = 1}^{N}\psi^{\dagger}(x)\psi(x) = 0
\end{align}
is fulfilled on the ``physical'' Hilbert space. States in the Hilbert space with $G(x)\neq 0$ are termed ``unphysical''. The Hamiltonian commutes with the Gauss's law generator $G(x)$ and, therefore, the dynamics is gauge invariant, meaning it connects physical states only to other physical states.

Following Ref.~\cite{Kogut1975}, the above continuum Hamiltonian is discretized onto a lattice with spacing $a$  
by putting matter with creation and annihilation operators $\psi_{n,p}^{\dagger}$, respectively, $\psi_{n,p}$ onto lattice sites labeled $n=1\dots N$ and electric fields $E_n$ on the links from site $n$ to $n+1$. 
A subtlety of the lattice formulation with respect to the continuum is that instead of the vector potential $A_n$, the lattice Hamiltonian is defined through the so-called link operator or parallel transporter $U_n = \exp(ieaA_n)$. The link operator describes the dynamical Peierls phase~\cite{Struck2012} picked up by a fermion moving from lattice site $n$ to $n+1$. Acting on a state, it raises the electric field by one unit.
The resulting Hamiltonian is 
\begin{align}
    H_{\text{KS}} = &\frac{a}{2}\sum_{n}(E_{n})^2 + \frac{ea\theta}{2\pi}\sum_n E_n + \frac{e^2a}{8\pi^2}\theta^2\notag\\ +& \sum_{n,p}(-1)^{n+p-1}m_{p}\psi^{\dagger}_{n,p}\psi_{n,p} \notag\\ -&\frac{i}{2a}\sum_{n,p} (f_{n,p}\psi^{\dagger}_{n,p}U_{n}\psi_{n+1,p}-\text{h.c.})\,.
 \label{eq:Hamiltonian_KS_SM}
\end{align}
The first term represents the electric field energy and the second the topological $\theta$ angle, equivalent to a background field. We also included a constant proportional to $\theta^2$ in order to be able to more clearly compare ground-state energies at different $\theta$ angles. The term in the second line represents the fermionic rest masses. 
Here, we use staggered fermions and chose to stagger the two flavors in opposite ways. Namely, the first flavor's particles (antiparticles) live on even (odd) sites, and opposite for the second flavor. This trick allows us to preserve a discrete chiral symmetry on the lattice, present in the continuum Schwinger model~\cite{Berruto1999, Dempsey2022}.
Finally, the third line represents the fermionic kinetic energy, which is coupled to the gauge field. 
For $f_{n,p} \equiv 1$, the system is under standard boundary conditions, while $f_{n,p} = -1$, if $n = L$ and $p = 2$, and $+1$ otherwise implements the flavor-twisted boundary conditions in our many-body lattice model.

On the lattice, states are considered physical if they obey the discretized Gauss's law $G_n\ket{\psi} = 0$, where 
\begin{align}
    \label{eq:discreteGaussLaw}
    G_n = E_n - E_{n-1} +1 -e\sum_p\psi^{\dagger}_{n,p}\psi_{n,p}.
\end{align}
The constant $1$ is added to the Gauss's law due to the particular staggering of the charges. 

In view of implementation in a numerical or quantum simulation, we further truncate the 
infinite Hilbert space of each gauge field. 
One common possibility is the Quantum Link Model~\cite{Chandrasekharan1997} representation, where the local Hilbert space on each link is given by that of a spin-$S$ object. 
In that formulation of quantum electrodynamics, the electric field operator $E_n$ on a link is replaced by the spin-$Z$ operator $eS_n^Z$; the link operator $U_n$ is replaced by $[S(S+1)]^{-1/2}S_n^+$ with matrix elements $\bra{m'}S_n^+\ket{m} = \delta_{m',m+1}\sqrt{1-m(m+1)/[S(S+1)]}$.
Here, instead, we consider the so-called truncated Schwinger model (TSM)~\cite{Zohar2012,Desaules2023,Desaules2023b}, where $U_n$ is replaced by the operator $\Tilde{S}_{n}^+$, whose matrix elements are given by  $\bra{m'}\tilde{S}_n^+\ket{m} := \delta_{m',m+1}$, up to a sharp cutoff at $\pm S$. 

The Hamiltonian of the resulting TSM reads for two flavors ($p=1,2$) and  including the $\theta$ angle
\begin{align}
    H_{\text{TSM}} = &\frac{e^2a}{2}\sum_{n}(S^z_{n})^2 + \frac{e^2a\theta}{2\pi}\sum_n S^z_n + \frac{e^2a}{8\pi^2}\theta^2\notag\\ +& \sum_{n,p}(-1)^{n+p-1}m_{p}\psi^{\dagger}_{n,p}\psi_{n,p} \notag\\ -&\frac{i}{2a}\sum_{n,p} (f_{n,p}\psi^{\dagger}_{n,p}\Tilde{S}^+_{n}\psi_{n+1,p}-\text{h.c.})\,.
 \label{eq:Hamiltonian_TSM_SM}
\end{align}
Both QLM and TSM formulations alter the gauge field operator algebra. Within the TSM, the gauge-field commutation relations are modified from $[E_n,U_m]=e \delta_{nm}U_m$ and $[U_n,U_m^\dagger]=0$ to
\begin{subequations}
\begin{align}
    [S^z_n,\Tilde{S}^{\pm}_m] = \pm \delta_{nm}\Tilde{S}_n^{\pm},\\
    [\Tilde{S}^+_n,\Tilde{S}^-_m] = 
    \delta_{nm}(\ket{S}\bra{S}-\ket{-S}\bra{-S}).
\label{eq:commutation_rel_TSM}
\end{align}   
\end{subequations}
Nonetheless, exact $U(1)$ gauge invariance is retained, as can be checked by noting that with the replacement $E_n\rightarrow eS_n^Z$; the Gauss's law generator given in Eq.~\eqref{eq:discreteGaussLaw} commutes with the TSM Hamiltonian given in Eq.~\eqref{eq:Hamiltonian_TSM_SM}. Importantly, despite the lattice discretization and electric-field truncation, the TSM thus preserves key properties of the Schwinger model such as confinement, chiral symmetry breaking, etc.; see, e.g.,~\cite{Chandrasekharan1999,Wiese2013,Banerjee2013,Pichler2016}.

Somewhat more quantitatively, Eq.~\eqref{eq:commutation_rel_TSM} differs from zero only at the cutoff levels of the electric field. Even though the commutators do not smoothly converge to those of lattice Schwinger model as for the QLM~\cite{Desaules2023b}, one may thus speculate that the TSM captures well the low-energy properties of the (untruncated) lattice Schwinger model, something that we indeed confirm below. 
To quantify the deviation from the lattice Schwinger model, we introduce 
\begin{align}
    \Delta U^2 =  \sum_n\langle[\Tilde{S}^+_n,\Tilde{S}^-_n]^2\rangle = \sum_n\langle P^{(-S)}_n + P^{(+S)}_n\rangle,
\label{eq:tsm_error}
\end{align}
where $P^{(m)} = \ket{m}\bra{m}$. This quantity essentially measures how strongly the link operator fails to be unitary, and is applicable in any dimension and with arbitrary number of flavors.

In the following, we present compelling evidence for the presence of fractons in the TSM already at small truncations ($S\geq 2$). Furthermore, we will show that for $S\geq 3$, the chiral condensate in the ground state of the TSM coincides with the semiclassical continuum prediction, with  deviations correlated with the strength of the quantitiy defined in Eq.~\eqref{eq:tsm_error}.

\section{Numerical results}Our numerical simulations are based on exact diagonalization (ED) [with system sizes up to $L = (4$ sites $+\:4$ links$)$ and truncation up to $S=2$] and tensor networks (TN) [with up to $L = (20$ sites $+\:20$ links$)$ and $S=4$], see Appendix~\ref{app:numerics} for details.

To obtain a direct evidence for fractons in the TSM, we study the lowest energy states as a function of the topological $\theta$ angle. For small nonzero fermionic mass, $m_p = m \neq 0$, continuum path-integral calculations predict the $N$ lowest energy levels of the $N$-flavor Schwinger model to oscillate as a function of the $\theta$ angle as~\cite{Misumi2019}
\begin{align}
    E_k(\theta) = -2m\exp{\bigg(-\frac{\pi}{N\mu eL}\bigg)}\cos{\bigg(\frac{\theta + 2\pi k}{N}\bigg)},
\label{eq:crossing_energies}
\end{align}
where $k\in \{0,\cdots,N-1\}$. 
That is, the fracton configurations become manifest through a $2\pi$ periodicity of $\theta/N$, rather than only $\theta$ as one is used to from the single-flavor Schwinger model.  
This fractional $\theta$ dependence is the precise signature we are looking for. 

In Fig.~\ref{fig:fractons}, we compare the three lowest energy levels of the TSM as a function of the topological $\theta$ angle for flavor-twisted (upper panel) and flavor-independent (lower panel)  boundary conditions. 
We observe a clear gap closing at the points $\theta = \pm \pi$, which suggests that ground and first excited states switch roles in a nonadiabatic fashion. To corroborate this result, we compute the fidelity susceptibility~\cite{Zanardi2007,Gu2010,Hauke2010,Wang2015} of the ground state. 
The fidelity susceptibility measures how drastically a quantum state changes as a parameter in the model is modified (see Appendix~\ref{app:fid_suscept}) and has become a useful tool to identify phase transitions without, e.g., \textit{a priori} knowledge of the order parameter.
As the deltalike peak shows, at $\theta = \pm \pi$ the properties of the ground state change rapidly, indicating a true level crossing. In accordance with Eq.~\eqref{eq:crossing_energies}, this implies a fractional $\theta$ dependence of the ground state.  
In contrast, for flavor-independent boundary conditions, the gap remains nonzero even at $\theta = \pm\pi$. Accordingly, the fidelity susceptibility shows only a broad peak at these points, indicating no drastic property change in the ground state, in accordance with an integer $\theta$ dependence. 
Remarkably, the fracton signature derived from perturbative continuum calculations in the small-$m$ limit persists in a wide range of values of Hamiltonian parameters, including significant rest mass, small system sizes, small gauge-field cutoffs, and large lattice spacing (parameters used in the figure: $m/e=0.4$, $L = (4$ sites $+\:4$ links$)$, $S=2$,  and $ea = 1$).

\begin{figure}[t!]
    \centering
    \includegraphics[width = 8.6cm]{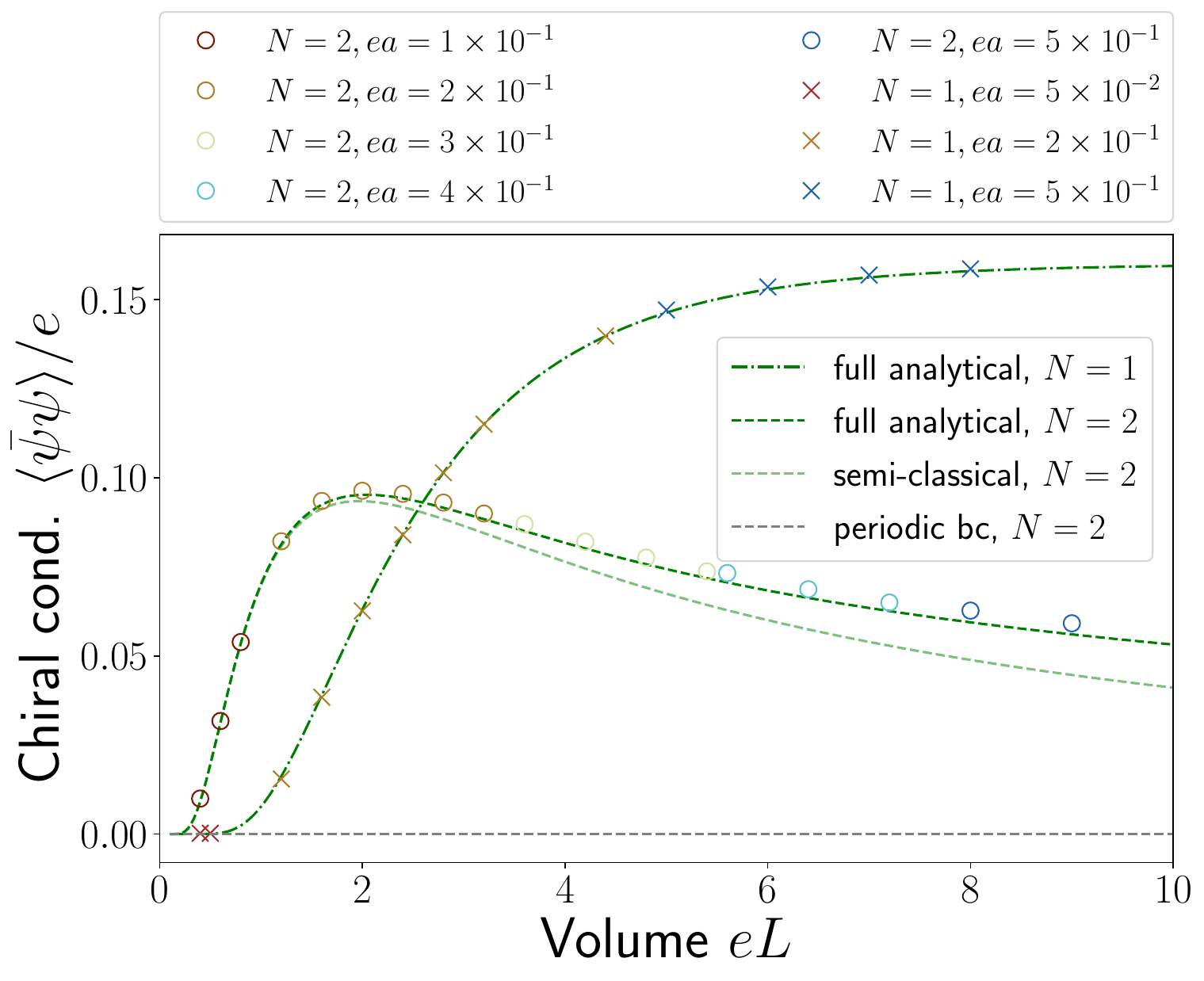}
    \caption{\textbf{Chiral condensate as a fingerprint of fractons in the zero-mass limit.} Throughout a large range of volumes, the TN expectation values of the chiral condensate for the TSM (cutoff $S = 3$) coincide with the analytic predictions from the continuum Schwinger model. 
    This result holds true also beyond the semiclassical approximation and is robust with respect to the lattice discretization and gauge-field cutoff. 
    For a single flavor, a chiral condensate appears due to the chiral anomaly . In the case of two flavors with standard boundary conditions, chiral symmetry is preserved in the ground state. With flavor-twisted boundary conditions, a chiral condensate is allowed and generated by the presence of fractons. 
    }
    \label{fig:improved_chiral_cond}
\end{figure}

\begin{figure}[t!]
    \centering
    \includegraphics[width = 8.6cm]{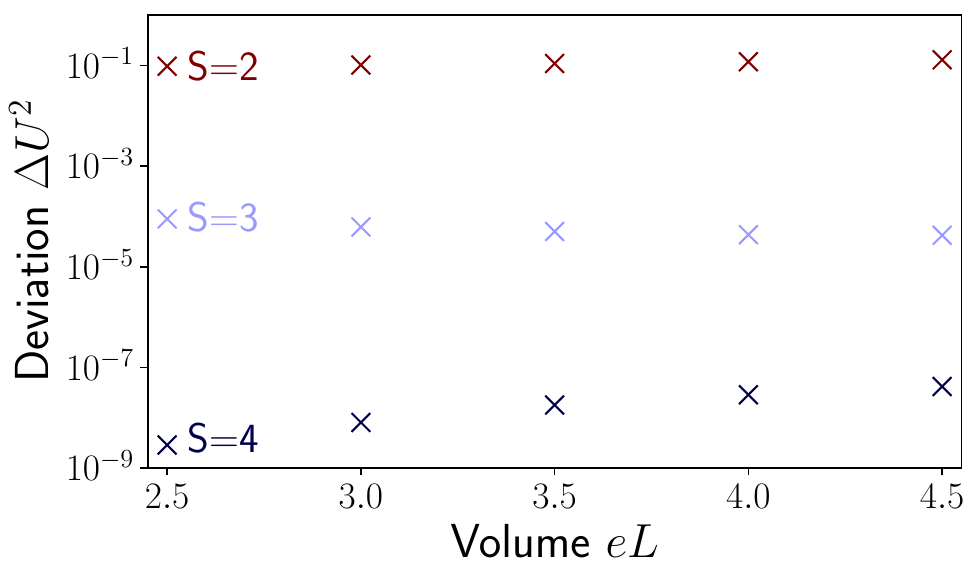}
\caption{
\textbf{Effect of finite gauge-field cutoff.}
For very small gauge-field truncations, the violation of the unitarity of the link operator, measured through Eq.~\eqref{eq:tsm_error}, is sizable, but already for $S = 3$, this deviation falls below a permille. 
The data suggests an exponential improvement with the cutoff size.
(Data for the ground state of the TSM from TN calculations for $m=0$ and $ea = 0.25$).}
    \label{fig:tsm_error_analysis}
\end{figure}

\section{Tensor network calculation of the chiral condensate}We obtain further evidence for the presence of fracton configurations  by investigating the chiral condensate, which on the lattice with staggered fermions becomes 
\begin{align}
    \langle\bar{\psi}\psi\rangle = \frac{1}{L}\sum_n(-1)^n\langle\psi_n^{\dagger}\psi_n\rangle
\end{align}
(here, we suppressed the flavor index).  Figure~\ref{fig:improved_chiral_cond} displays the main results of our TN simulations, for the range of physical volume $eL\in[0.4,9.0]$, lattice spacings $ea\in [0.1,0.5]$, and spin truncation $S=3$. 
As the flavor-twisted boundary conditions induce chiral symmetry breaking, a finite volume can support a nonvanishing chiral condensate. 
Quite surprisingly, even for the coarse truncation of $S=3$, we encounter throughout the entire range of volumes considered a quantitative agreement between the lattice calculations and the analytical predictions for the continuum Schwinger model~\cite{Shifman1994}.  
The insignificance of lattice artifacts on the chiral condensate is quite remarkable, considering we do not take the continuum limit~\cite{Banuls2016} and that lattice spacings used are as large as $ea = 0.5$. In the next section, we will devise a strategy for quantum simulation of the fracton physics on a trapped-ion qudit quantum device. In that setting, one can think of choosing improved Hamiltonians~\cite{Carena2022,Spitz2019} for even faster convergence of the chiral condensate results to the continuum limit. Such improvements could even often come with no overhead in terms of measurements count and therefore deliver more accurate results almost for free. For the purpose of this work, however, we do not need to take extra care of finite lattice spacing scaling, as rather coarse lattices already give excellent agreement with continuum predictions.

In Appendix~\ref{app:bond_dim_and_lattice_spacing}, we discuss the influence of the finite bond dimension on the MPS results and we show that the numerically accessible system sizes permit one to further improve the accuracy by extrapolations to vanishing lattice spacing at fixed volume.

The same analysis as above can be made for the single-flavor Schwinger model. Again, the lattice results with $S = 3$ and $ea \leq 0.5$ already essentially coincide with the continuum prediction. 
The single-flavor Schwinger model explicitly breaks chiral symmetry, leading to convergence of the chiral condensate to a nonzero value at $eL\rightarrow\infty$. 
In contrast, for the two-flavor model with standard boundary conditions, chiral symmetry is preserved. Flavor-twisted boundary conditions break this symmetry weakly and induce the formation of a chiral condensate, but its strength falls off with $L$, highlighting the fact that the effect of boundary conditions disappears for infinite volume. As considerations based on the path integral show [Eq.~\eqref{eq:chiralcond_fromZ}], the presence of a chiral condensate at vanishing rest mass reveals the contributions due to configurations with fractional topological charge.  




Finally, Eq.~\eqref{eq:tsm_error} allows us to monitor finite cutoff errors in the TSM. In Fig.~\ref{fig:tsm_error_analysis}, we show $\Delta U^2$ in a range of lattice volumes for cutoffs $S = 2,3,4$. Already for $S=3$, the occupation of the cutoff levels of the gauge fields is below a permille. The data suggests a suppression of the error that is exponential in the cutoff, thus leading to a rapid restoration of the commutation relations of the Schwinger model. 

\section{Quantum simulation}
Quantum simulators are rapidly evolving to access ever more complex gauge-theory phenomena~\cite{Zohar2016,Banuls2020,aidelsburger2022cold,osti_1892238,Humble2022vtm,Bauer2022hpo,beck2023quantum,DiMeglio2024,halimeh2023coldatom}, since recently also including the multiflavor Schwinger model \cite{funcke2022exploring}. While the ultimate goal is to perform large-scale simulations of non-Abelian gauge-theory models and tackle the mysteries of quantum chromodynamics, studying intricate physical phenomena as a consequence of gauge symmetries, such as fracton signatures, are an important intermediate stepping stone for demonstrating the capabilities of quantum simulation platforms. In this section, we detail a variational protocol based only on existing technology to provide a direct demonstration of fractons in an experimental setting.

\subsection{Local qubit-qudit encoding of the two-flavor TSM Hamiltonian}

To perform quantum simulations in the NISQ era, it is essential to minimize the experimental overhead. Therefore, we make use of the redundancy in the TSM model due to the Gauss's law, by extending the methods of Refs.~\cite{Zohar2019,Popov2024} to the multiflavor case. Specifically, by formulating a unitary transformation and applying it on the two-flavor TSM Hamiltonian, we show how to map a system of $L = (n$ sites $+\:n$ links$)$ onto a system of $n$ qubits and $n$ qudits, saving, in total, $n$ qubits with respect to the naive encoding. 

The first step toward the qubit-qudit Hamiltonian consists of performing a Jordan--Wigner (JW) transformation~\cite{Jordan1928}, in which the fermions are replaced by spin-1/2 particles, 
\begin{align}
    \psi_{n,1} &\rightarrow \prod_{i<n}\sigma^z_{i,1}\sigma^-_{n,1},\notag\\
    \psi_{n,2} &\rightarrow \mathcal{P}_1\prod_{i<n}\sigma^z_{i,2}\sigma^-_{n,2},        
\end{align}
where $\mathcal{P}_1$ is the parity of the flavor$-1$ fermions (at half filling, $\mathcal{P}_1 = +1$).
The JW transformation preserves the locality of the Hamiltonian in one spatial dimension. (In higher dimensions, one could exploit Gauss's law to map the fermions to hard-core bosons while maintaining locality of the model~\cite{Zohar2018}).

We further use the Gauss's law in order to unitarily eliminate one of the two spins per lattice site. The key observation is that the Gauss's law permits one to keep track of the total number of particles present on each site $n$ by evaluating the divergence of the electric field $S^z_n - S^z_{n-1}$. Therefore, knowing the divergence and the state of the first flavor directly gives us information about the second flavor. One can then formulate a unitary transformation on the physical Hilbert space that decouples the spins representing the second flavor from the Hamiltonian and maps them to their vacuum state. The unitary that achieves this transformation is 
\begin{align}
    \mathcal{U} = \prod_{n}(\sigma^X_{n,2})^{S^z_n-S^z_{n-1}+\frac{1}{2}(1-\sigma^z_{n,1})}.
\end{align}
The transformed TSM Hamiltonian reads
\begin{align}
    \tilde{H}_\mathrm{TSM} =& \mathcal{U}H_\mathrm{TSM}\mathcal{U}^{\dagger}\notag\\ = &\frac{e^2a}{2}\sum_{n}(S^z_{n})^2 + \frac{e^2a\theta}{2\pi}\sum_n S^z_n + \frac{e^2a}{8\pi^2}\theta^2\notag\\ +& \frac{m_1-m_2}{2}\sum_{n}(-1)^{n}\sigma^z_n + m_2\sum_{n}S^z_n\notag\\ +&\frac{1}{2a}\sum_{n} (\sigma^+_n\tilde{S}^+_{n}\sigma^-_{n+1}+\text{h.c.})\notag\\+&\frac{1}{2a}\sum_{n} (P^1_n\tilde{S}^+_{n}P^1_{n+1}+\text{h.c.})\,,
\label{eq:cost_function}
\end{align}
where we have suppressed the flavor index of the remaining spin-1/2 particle and $P^1_n$ denotes the projector of the operator $S^z_n-S^z_{n-1}+\frac{1}{2}(1-\sigma^z_n)$ onto the eigenstate with eigenvalue $+1$. The content on the links has not changed, but we have reduced the number of spins per matter site from two to one. Precisely this is the Hamiltonian that will serve as our cost function for the variational algorithm we employ below.

\subsection{Numerical tests for a variational state preparation on a trapped-ion qudit quantum platform}

\begin{figure}[t!]
    \centering
    \includegraphics[width = 8.6cm]{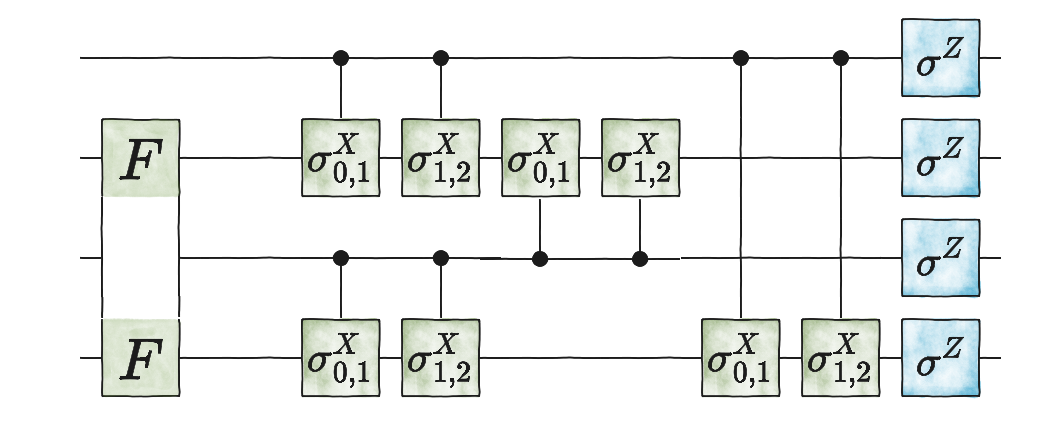}
\caption{
\textbf{Individual layer of the variational circuit used for ground- and excited-state preparation.} After a sequence of entangling two-body gates between the qubits representing the sites and the qutrits representing the links of the lattice, a sequence of single-qubit and single-qutrit gates is applied. Each quantum gate is parametrized by a variational parameter. The software used for drawing the quantum circuits is taken from~\cite{drawio_library}.}
    \label{fig:variational_circuit}
\end{figure}

\begin{figure}[t!]
    \centering
    \includegraphics[width = 8.6cm]{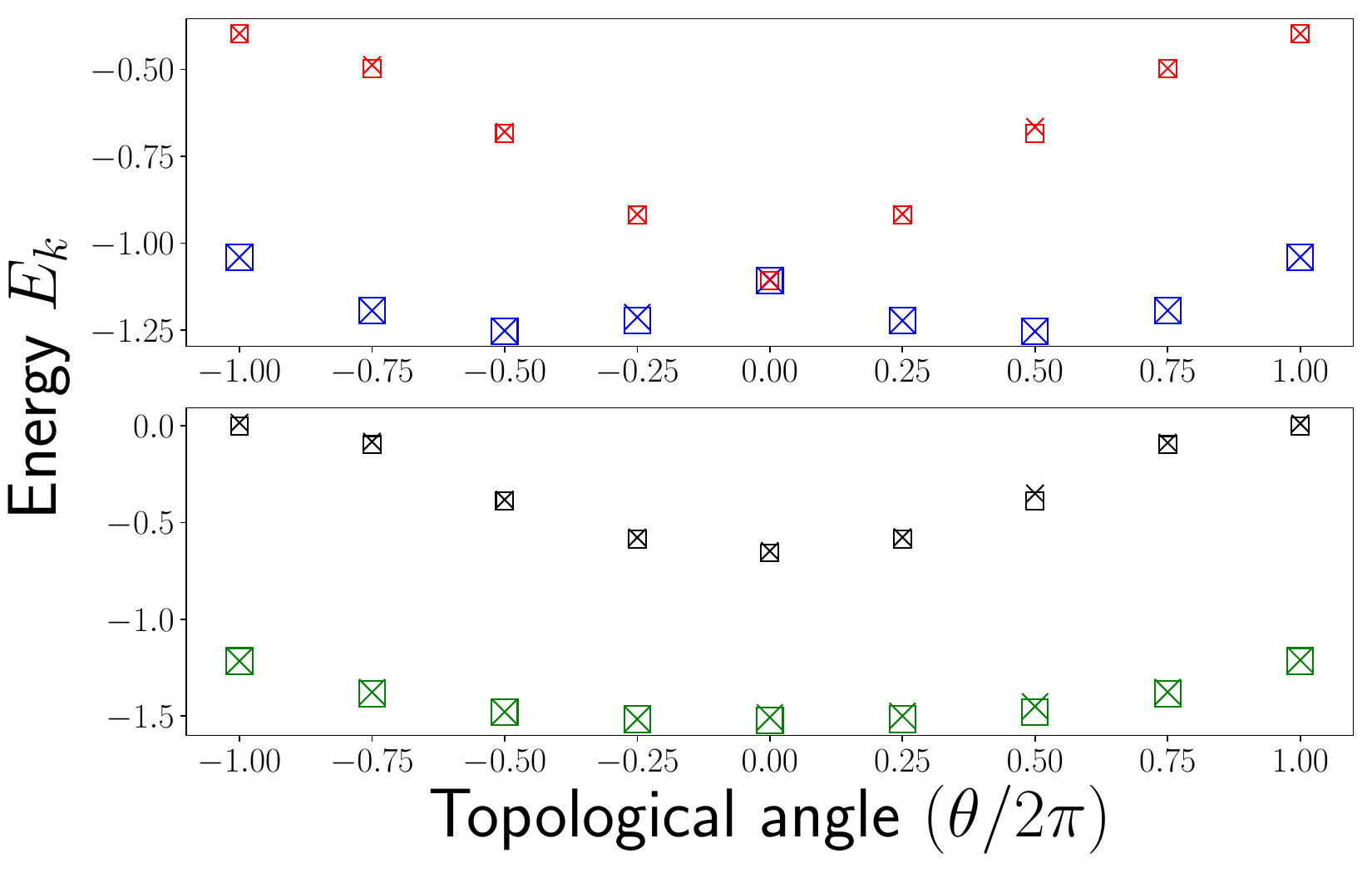}
\caption{
\textbf{Results from variational state preparation for a system of two sites and two links.} Using a quantum circuit with three layers as shown in FIG.~\ref{fig:variational_circuit} and applying the VarQITE algorithm, we prepare the ground and excited state of the model with periodic (green crosses) and with flavor-twisted (blue crosses) boundary conditions. Even for only three layers, fidelities of $>99\%$ with the ground state and $>90\%$ with the first excited state (black and red crosses, respectively) are achieved. The found energy values agree extremely well with the numerical values from exact diagonalization (colored squares). The gap closing for flavor-twisted boundary conditions versus the open gap for periodic boundary conditions can be clearly identified (note the $\pi$ shift of the $\theta$ angle with respect to FIG.~\ref{fig:fractons}, which can be performed by changing the sing of the fermion mass due to the chiral anomaly).
The slight asymmetry with respect to $\theta\to -\theta$ in the energy value is due to the random choice of the starting point of the variational minimization procedure.}
    \label{fig:variational_results}
\end{figure}

To provide further evidence for the near-term implementability of our proposed setup, we present numerical simulations of ground and excited state preparation using the variational quantum imaginary time evolution (VarQITE) algorithm~\cite{Yuan2019}. These simulations were conducted on a proof-of-principle system consisting of (2 sites + 2 links), further corroborating the feasibility of our approach in small-scale scenarios. 

As we will show in the following, a spin truncation of $S = 1$ already allows for observing the characteristically different behavior for periodic and for flavor-twisted boundary conditions. In particular, the gap closing at a topological angle of $\theta = \pi$ for flavor-twisted boundary conditions, evidencing the altered periodicity of the ground state energy as a function of $\theta$, can be clearly observed in that spin truncation. 

The variational circuit we use here 
is composed of gates that are easy to implement on state-of-the-art trapped-ion qudit platforms~\cite{Ringbauer2022}. The circuit is designed in a layered structure, where each layer consists of entangling two-body and single-body operations as depicted in FIG.~\ref{fig:variational_circuit}. In order to create entanglement between the qubits that represent the matter sites and the qudits representing the gauge links, we employ the so-called controlled rotation (CROT) gates:
\begin{align}
    \mathrm{CROT}_{k,ij}(\lambda) = \big(\mathbb{1}-\ket{k}\bra{k}\big)\otimes\mathbb{1} + \ket{k}\bra{k}\otimes\exp\big(-i\lambda\sigma_X^{ij}\big).
\end{align}
In the expression above, $\sigma_X^{ij}$ is the two-level $\sigma$ matrix, defined as follows
\begin{align}
    (\sigma^{i,j}_X)_{m,n} &= \begin{cases}
    1\,, \quad \text{if }  (m,n) = (i,j) \text{ or } (j,i)  \\
    0\,, \quad \text{otherwise.}
    \end{cases}\\
\label{eq:two_level_pauli_matrices}
\notag
\end{align}

In addition, we entangle the qubits on the sites with the so-called flip-flop gate
\begin{align}
    \mathrm{FF}_{nm}(\lambda) = \exp\big[-i\lambda(\sigma^{+}_n\sigma^{-}_m + \sigma^{-}_n\sigma^{+}_m) \big].
\end{align}
Here, $\sigma^{\pm}_n$ is the raising (lowering) operator on the $n$th qubit. At the end of each layer, single qubit and qudit rotations are applied. Specifically, we find that simple $\sigma^Z $rotations are sufficient
\begin{align}
    U_n(\lambda) = \exp\big(-i\lambda\sigma^Z_n\big).
\end{align}
In the equation above, $\sigma^Z_n$ is the Pauli-$Z$ matrix in case of a qubit and the spin-1 $Z$-operator in case of a qutrit.

To find a variational approximation to the ground state, we employ the usual VarQITE protocol, where the initial state corresponds to a random initialization of the quantum circuit with two layers as in FIG.~\ref{fig:variational_circuit} on top of the product state $\ket{\Psi_0} = \ket{1011}$ (nomenclature: qubit 0, qubit 1, qudit 0, qudit 1). Subsequently, evaluating the cost function given by the Hamiltonian in Eq.~\eqref{eq:cost_function} and optimizing the parameters of the various gates iteratively until convergence is reached delivers the desired approximation of the ground state.
To prepare the first excited state, we add a penalty term to the Hamiltonian with respect to the (variationally found) ground state. Notably, all four relevant states---ground and first excited states for periodic and for flavor-twisted boundary conditions---can be prepared with high fidelity using the same quantum circuit, for all values of $\theta$ used. The only difference is the gate parameters to which the algorithm converges.

In FIG.~\ref{fig:variational_results}, we display the results of the variational state preparation in comparison to exact numerical results. Strikingly, already for $S=1$ and only three layers the variational state preparation allows for clearly distinguishing the distinct behavior of the gap for periodic and for flavor-twisted boundary conditions. 
The above algorithm uses only standard gates and the number of entangling two-qudit gates for the circuit with two layers is $N_{ent} = 27$, well within the coherence time limit of current trapped-ion devices. The actual bottleneck for an accurate quantum simulation on the trapped-ion device is the fidelity of an individual entangling gate. In case we assume this fidelity to be $99\%$, the total fidelity of a quantum circuit with $N_{ent} = 27$ entangling gates as in Fig.~\ref{fig:variational_circuit} will be roughly $(99\%)^{27}\approx 76\%$~\cite{Meth2023}. Moreover, since already very coarse truncations of the electric field are sufficient, the dimensionality of the qudits can be as small as $d = 5$ ($S=2$)  for qualitative signatures and as small as $d = 7$ ($S=3$) for quantitative agreement with continuum results. The fractionalization can thus be observed with existing technology. 

At last, we would like to draw to the attention of the reader the existence of alternative variational simulation algorithms, such as AVQITE~\cite{Gomes2021} and SC-ADAPT-VQE~\cite{farrell2024quantum}. Those algorithms have been shown, in some cases, to perform even better than the VarQITE algorithm used here, where the second one even enabled large-scale simulations for simple gauge theory models. Further improvements may be achieved by employing preoptimization techniques~\cite{khan2023,Mullinax2025}, where an initial optimization is performed on a classical device and the set of variational parameters is given as a warm start to the quantum device. While the effort for quantum simulation of gauge theory models, such as the one we presented in our work, could definitely benefit from such sophisticated algorithms, this is not the focus of our work here. Instead, our work shows that fracton signatures are so robust that even simple algorithms such as VarQITE can reveal them on a quantum device.

\section{Conclusions}In this article, we have investigated fractional gauge field configurations---a version of instantons with a fractional topological charge---and how they contribute to the buildup of a chiral condensate in the multiflavor Schwinger model. Going beyond semiclassical approximations~\cite{Shifman1994}, we have demonstrated the presence of fractons in a strongly correlated many-body lattice model for various system sizes and in a range of values for the mass and the coupling constant. We have identified the smallest spin truncation ($S = 3$), for which the results for the chiral condensate in the zero-mass limit agree quantitatively with the predictions from the continuum theory. The fingerprints of fractons are thus remarkably robust against nonperturbative effects and discretization and truncation artifacts. 
As we have shown through numerical benchmark simulations of a VarQITE algorithm, this robustness makes it possible to experimentally observe fracton signatures in already existing quantum simulators.    


In this work, we have restricted our analysis to two flavors coupled to a $U(1)$ gauge field. Immediate extensions could investigate large number of fermionic flavors---a test bed for probing phenomena like chiral phase transition in QCD~\cite{PhysRevD.29.338,Aoki2006,wu2024exploring}. 
Even though we have revealed signatures of gauge configurations with fractional topological charge in nonperturbative settings, understanding their origin beyond semiclassical arguments remains a long-standing problem. A future research direction could be to develop a characterization in terms of topological invariants valid for an interacting many-body system with local gauge symmetry, in analogy to a similar effort in condensed matter systems~\cite{PhysRevX.2.031008,Elben2020,PhysRevB.107.125161}.

\section{Acknowledgments} We acknowledge useful discussions with Martin Ringbauer, Matteo Wauters, Emanuele Tirrito, and Niccolò Baldelli. 

ICFO group acknowledges support from: ERC AdG NOQIA; MCIN/AEI (PGC2018-0910.13039/501100011033, CEX2019-000910-S/10.13039/501100011033, Plan National FIDEUA PID2019-106901GB-I00, Plan National STAMEENA PID2022-139099NB-I00 project funded by MCIN/AEI/10.13039/501100011033 and by the “European Union NextGenerationEU/PRTR" (PRTR-C17.I1), FPI); QUANTERA MAQS PCI2019-111828-2); QUANTERA DYNAMITE PCI2022-132919 (QuantERA II Programme co-funded by European Union’s Horizon 2020 program under Grant Agreement No 101017733), Ministry for Digital Transformation and of Civil Service of the Spanish Government through the QUANTUM ENIA project call - Quantum Spain project, and by the European Union through the Recovery, Transformation and Resilience Plan - NextGenerationEU within the framework of the Digital Spain 2026 Agenda; Fundació Cellex; Fundació Mir-Puig; Generalitat de Catalunya (European Social Fund FEDER and CERCA program, AGAUR Grant No. 2021 SGR 01452, QuantumCAT \ U16-011424, co-funded by ERDF Operational Program of Catalonia 2014-2020); Barcelona Supercomputing Center MareNostrum (FI-2023-1-0013); EU Quantum Flagship (PASQuanS2.1, 101113690, funded by the European Union. Views and opinions expressed are however those of the author(s) only and do not necessarily reflect those of the European Union or the European Commission. Neither the European Union nor the granting authority can be held responsible for them); EU Horizon 2020 FET-OPEN OPTOlogic (Grant No 899794); EU Horizon Europe Program (Grant Agreement 101080086 — NeQST), results incorporated in this standard have received funding from the European Innovation Council and SMEs Executive Agency under the European Union’s Horizon Europe programme), ICFO Internal “QuantumGaudi” project; European Union’s Horizon 2020 program under the Marie Sklodowska-Curie grant agreement No 847648; “La Caixa” Junior Leaders fellowships, La Caixa” Foundation (ID 100010434): CF/BQ/PR23/11980043. Views and opinions expressed are, however, those of the author(s) only and do not necessarily reflect those of the European Union, European Commission, European Climate, Infrastructure and Environment Executive Agency (CINEA), or any other granting authority. Neither the European Union nor any granting authority can be held responsible for them.

P.P.P.~acknowledges also support from the “Secretaria d’Universitats i Recerca del Departament de Recerca i Universitats de la Generalitat de Catalunya” under grant 2024 FI-3 00390, as well as the European Social Fund Plus.

P.H.~has received funding from the European Union’s Horizon Europe research and innovation programme under grant agreement No 101080086 NeQST and the Italian Ministry of University and Research (MUR) through the FARE grant for the project DAVNE (Grant R20PEX7Y3A). 
This project was funded within the QuantERA II Programme that has received funding from the European Union’s Horizon 2020 research and innovation programme under Grant Agreement No 101017733, by the European Union under NextGenerationEU, PRIN 2022 Prot. n. 2022ATM8FY (CUP: E53D23002240006), 
by the European Union under NextGenerationEU via the ICSC – Centro Nazionale di Ricerca in HPC, Big Data and Quantum Computing, by the Provincia Autonoma di Trento, and Q@TN, the joint lab between University of Trento, FBK—Fondazione Bruno Kessler, INFN—National Institute for Nuclear Physics, and CNR—National Research Council.
Views and opinions expressed are however those of the author(s) only and do not necessarily reflect those of the European Union, The European Research Executive Agency, or the European Commission.  Neither the European Union nor the granting authority can be held responsible for them.


E.Z.\ acknowledges  the support of the Israel Science Foundation (grant No. 523/20). 

\section{Data availability} 
The data set used in this paper is available on Zenodo~\cite{popov_2025_zenodo}.

\bibliography{ref}
\appendix
\section{Lagrange formulation, path integral of the Schwinger model, and fractional gauge configurations}
\label{app:path_integral}
In this section, we briefly summarize the main notation useful for understanding the path integral formulation of QED in $(1+1)$ dimensions. We also explain how fractons arise in the Euclidean path integral of this theory. The Lagrange density of the multiflavor Schwinger model, including the topological $\theta$ angle, on a torus $\mathbb{T} = \mathbb{S}_{\beta}\times \mathbb{S}_L$, where $\beta = 1/T$ is the inverse temperature and $L$ is the spatial volume, reads
\begin{align}
\mathcal{L}_{\theta}(\mathbf{x}) = &\frac{ie\theta}{4\pi}\epsilon_{\mu\nu}F_{\mu\nu}(\mathbf{x})-\frac{1}{4}F_{\mu\nu}(\mathbf{x})^2 +\notag\\&i\sum^{N}_{p=1}\bar{\psi}_p(\mathbf{x})\slashed{D}(\mathbf{x})\psi_p(\mathbf{x}) - \sum^{N}_{p=1}m_p\bar{\psi}_p(\mathbf{x}) \psi_p(\mathbf{x})\,,
\label{eq:lagrangian}
\end{align}
where $\mathbf{x} = (\tau,x)$ with $\tau$ being the Euclidean time. Here, we use Euclidean signature with $\epsilon_{\mu\nu}$ being the total antisymmetric tensor in two dimensions, $F_{\mu\nu} = \partial_{\mu}A_{\nu}-\partial_{\nu}A_{\mu}$ the field strength tensor of the $U(1)$ gauge field $A_{\mu}$, $\psi_p$ the two-component Dirac field of the $p$'th flavor with bare mass $m_p$, $e$ the coupling constant, and $N$ the number of flavors. Without loss of generality, the $\theta$ angle can be restricted to values in the interval $\theta\in[0,2\pi]$. The Euclidean action is defined as the integral over the volume of the Lagrange density $S_{\theta} = \int_{\mathbb{T}}d^2x\mathcal{L}_{\theta}(\mathbf{x})$.

The object of central interest is the Euclidean path integral (or the partition function) of the Schwinger model, which is defined as an integral over all configurations for $\psi$ and $A_{\mu}$ 
\begin{align}
    Z = \int\mathcal{D}A_{\mu}\mathcal{D}\psi\mathcal{D}\bar{\psi}e^{-S_{\theta}[\bar{\psi},\psi,A_{\mu}]}.
\end{align}
In the differential of the integral, a product over the fermion flavors is implied. 
Via the anomaly, the $\theta-$term in the action can be traded for a summation over partition functions $Z_n$ deriving from sectors with distinct topological charge $\nu_2 = n$, i.e.,  
\begin{align}
    \label{eq:sumZn}
    Z = \sum_n Z_n = \sum_ne^{-in\theta}\int_n \mathcal{D}A_{\mu}\mathcal{D}\psi\mathcal{D}\bar{\psi}e^{-S[\bar{\psi},\psi,A_{\mu}]}\,.
\end{align}
Here, $S = \int_{\mathbb{T}}d^2x\mathcal{L}_{\theta=0}(\mathbf{x})$ and we use the index $n$ to denote integration only over gauge-field configurations with specified winding corresponding to topological charge $n$. 

The topology in the gauge sector is determined by the boundary conditions. 
Gauge-field configurations with integer topological charge $v_2 = n\in \mathbb{Z}$, or ``instantons'', satisfy standard boundary conditions (sbc) in time direction
\begin{align}
    A_0(\tau + T,x) &= A_0(\tau,x),\notag\\
    A_1(\tau + T,x) &= A_1(\tau,x) + \frac{2\pi n}{eL},\notag\\
    \psi_p(\tau + T,x) &= e^{-2\pi n i x/L}\psi_p(\tau,x)\,.
\label{eq:pbc_euclidean}
\end{align}
However, this is not the only possible choice for boundary conditions of the gauge fields. In the presence of multiple fermionic flavors, the ``large gauge transformation'' allows for boundary conditions that correspond to a fractional topological charge $\nu_2 = m/N,\: m\in \mathbb{Z}$:
\begin{align}
    A_0(\tau + T,x) &= A_0(\tau,x),\notag\\
    A_1(\tau + T,x) &= A_1(\tau,x) + \frac{2\pi m}{NeL},\notag\\
    \psi_p(\tau + T,x) &= \psi_{p+1}(\tau,x)\text{ for } p\in\{1,\cdots,N-1\},\notag\\
    \psi_N(\tau + T,x) &= e^{-2\pi m i x/L}\psi_1(\tau,x)\,.
\label{eq:tbc_euclidean}
\end{align}
The gauge field configurations that satisfy Eq.~\eqref{eq:tbc_euclidean} are called fractons. 
By including them in the Euclidean path integral, the summation in Eq.~\eqref{eq:sumZn} is taken over $n=m/N$, with $m\in \mathbb{Z}$. In contrast, for the single-flavor Schwinger model, the summation is directly over integer $n\in \mathbb{Z}$. 
As Eq.~\eqref{eq:sumZn} shows, the periodicity with respect to the $\theta$ angle increases accordingly to $N2\pi$. 
The existence of fractons can thus modify the properties of the ground state in a profound way.

\section{Detecting fractons via the chiral condensate in the presence of flavor-twisted boundary conditions}
\label{app:detecting}
The fractons in the multi-flavor Schwinger model can be revealed by detecting their contribution to observables like the chiral condensate. When flavor-independent (``standard'') boundary conditions in space direction are imposed on the fermions, i.e., $\psi_p(x+L) = e^{i\alpha}\psi_p(x) \:\forall p\in\{1,\cdots,N\}$, potentially with a flavor-independent phase $\alpha$, the system has a $SU(N)_L\otimes SU(N)_R$ flavor symmetry in the zero mass limit, prohibiting the generation of a chiral condensate due to the Mermin--Wagner--Coleman theorem~\cite{Coleman1973}. 
However, in case the 
phase $\alpha$ is flavor dependent, e.g., $\alpha_p = 2\pi p/N$, the chiral symmetry is explicitly broken and chiral condensation can be allowed even for vanishing rest masses \cite{Shifman1994}.

semiclassical analytics for small volumes ($eL\ll 1$) as well as exact path-integral calculations at arbitrary volumes~\cite{Shifman1994} (both at vanishing rest mass),  show that such flavor-twisted boundary conditions (fbc) indeed result in a nonvanishing chiral condensate. The main ingredient of this analysis is a new symmetry that the ground state obeys under fbc, induced by the ``fractional'' transformation
\begin{align}
    \label{eq:symmetry}
    A_1(x) &\stackrel{\mathrm{fbc}}{\longrightarrow} \tilde{\mathcal{S}}[A_1(x)] =  A_1(x) + \frac{2\pi}{NeL},\notag\\
    \psi_p(x)&\stackrel{\mathrm{fbc}}{\longrightarrow} \tilde{\mathcal{S}}[\psi_p(x)] =  e^{-i \frac{2\pi x}{NL}} \psi_{p+1}(x),
\end{align}
in contrast to the symmetry transformation for standard boundary conditions
\begin{align}
    A_1(x) &\stackrel{\mathrm{sbc}}{\longrightarrow} \mathcal{S}[A_1(x)] =  A_1(x) + \frac{2\pi}{eL},\,\,\,\notag\\
    \psi_p(x) &\stackrel{\mathrm{sbc}}{\longrightarrow} \mathcal{S}[\psi_p(x)] =  e^{-i \frac{2\pi x}{L}} \psi_{p}(x).
\end{align}

The ``fractional'' transformation is a combination of broken chiral symmetry and forbidden ``large gauge transformation'' and has a period $1/N$ times smaller than for sbc. The existence of this new symmetry requires one to impose a new superselection rule, one in which the length of a noncontractable circle in the space of gauge fields is reduced by the number of flavors $N$. On the level of the path integral, this symmetry leads to the increased $\theta$-periodicity discussed in the previous section. 

Exploiting this symmetry, path-integral calculations that are possible in the limit of vanishing rest mass lead to the  analytic predictions for the $L$-dependent chiral condensate~\cite{Shifman1994}, as given in the main text: 
\begin{align}
\langle\bar{\psi}\psi\rangle = \sqrt{\frac{\mu e^{\gamma}}{16\pi L}}e^{-I(L,\mu)/2},
\end{align}
with $\mu^2 = Ne^2/\pi$ the photon mass, $\gamma$ Euler's constant, and
\begin{align}
    I = \int_0^{\infty} \frac{d\omega}{\sqrt{\omega^2+\mu^2}}\Bigg(\coth{\frac{L\sqrt{\omega^2+\mu^2}}{2}}-1\Bigg)\,.
\end{align}

\section{Numerical implementation}
\label{app:numerics}
Our numerical simulations are based on two techniques: exact diagonalization (ED) using the QuSpin package~\cite{Weinberg2019} in Python, employed for system sizes up to $L = (4$ sites $+\:4$ links$)$, with a spin $1/2$ on each site and up to spin $S=3$ on each link; and tensor network (TN) calculations using the DMRG method in the library ITensors~\cite{itensor} in Julia,  with system sizes up to $L = (20$ sites $+\:20$ links$)$ and similar spin sizes as for ED. For the TN calculations, the maximal bond dimension used in the MPS representation of the variational state is $\chi_{\mathrm{max}} = 400$, which we found sufficient for obtaining converged results, see Appendix~\ref{app:bond_dim_and_lattice_spacing} for discussion on convergence. 
In our numerical simulations, we implement the dimensionless version~\cite{Banuls2016} of the Hamiltonian of Eq.~(5) - $H \to H/e^2a$.

A main observable that serves as a signature for the presense of fractons is, as elaborated in the main text, the chiral condensate. Due to the degeneracy of the ground state that occurs for zero fermionic mass and also at topological angle $\theta = \pi$, the chiral condensate has to be extracted from a particularly chosen superposition of the two ground states found numerically. Denoting the two numerical ground states with $\ket{\Psi_1}$ and $\ket{\Psi_2}$, respectively, we define the superposition $\ket{\Psi_{\alpha}} = \cos(\alpha)\ket{\Psi_1} + \sin(\alpha)\ket{\Psi_2}$. Then, the chiral condensate plotted in FIG.~2 in the main text is $\langle\bar{\psi}\psi\rangle = \text{max}_{\alpha}\bra{\Psi_{\alpha}}\bar{\psi}\psi\ket{\Psi_{\alpha}}$.

\section{Finite bond dimension and lattice spacing effects}
\label{app:bond_dim_and_lattice_spacing}

\begin{figure}[t!]
    \centering
    \includegraphics[width = 8.6cm]{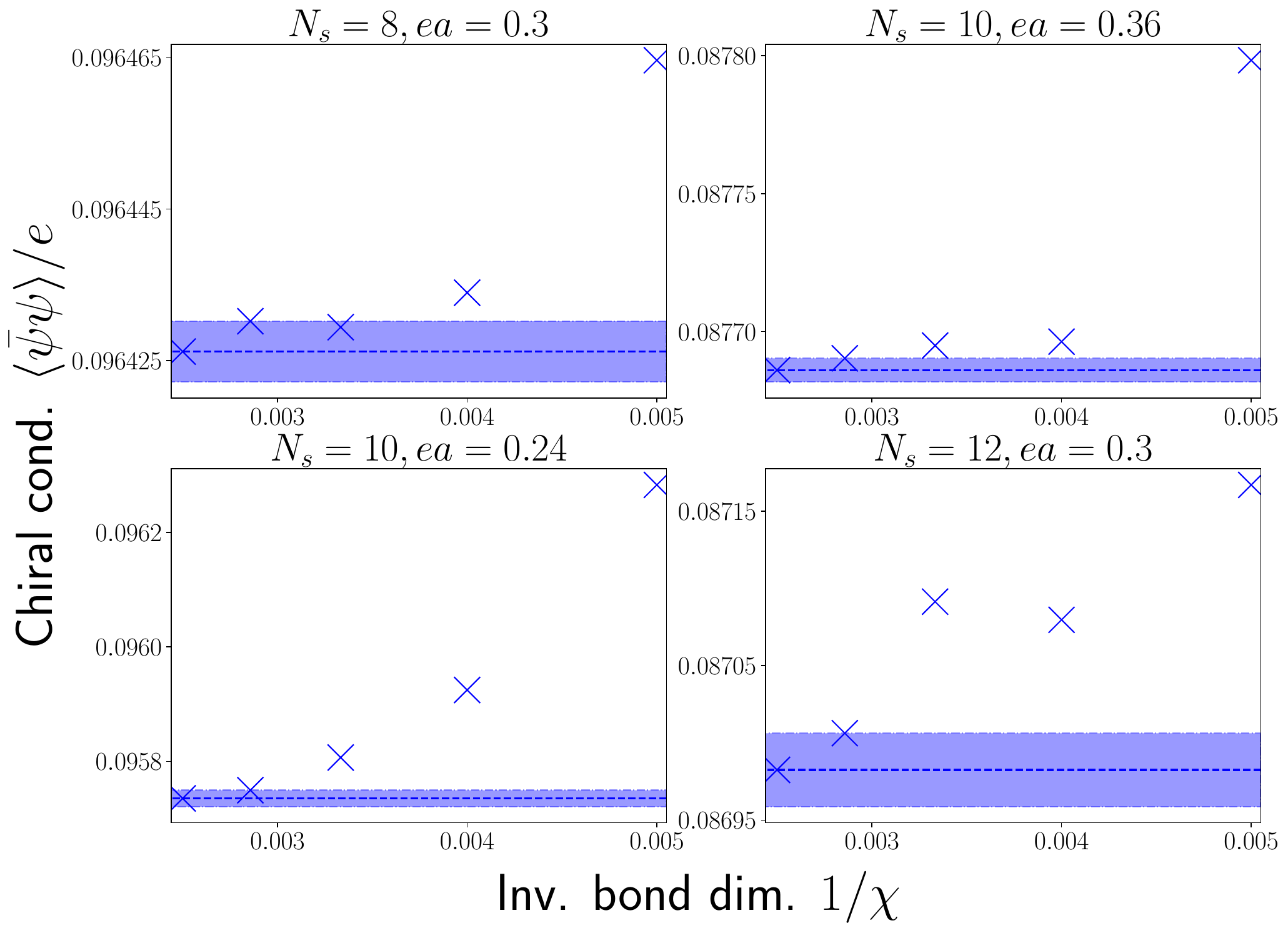}
\caption{
\textbf{Extrapolation with bond dimension.} The numerical value of the chiral condensate is plotted as a function of the inverse maximal bond dimension used in the DMRG calculation, for fixed lattice size $N_s$ and lattice spacing $ea$. As done in Ref.~\cite{Banuls2016}, the value for the chiral condensate at infinite bond dimension (dashed line) is taken as the one for largest calculated bond dimension (here, $\chi_{\text{max}} = 400$) and the error bars are the difference between the values for $\chi = 400$ and $\chi = 350$. Lower values of the bond dimension ensure that the true value for $\chi = \infty$ lies in the uncertainty interval (shaded area). The determined values and error bars are $0.096426(4)$ for $N_s = 8,ea = 0.3$, $0.095735(14)$ for $N_s = 10,ea = 0.24$, $0.087686(4)$ for $N_s = 10,ea = 0.36$ and $0.086982(23)$ for $N_s = 12,ea = 0.3$, respectively. The maximal relative error for these values is on the order of 2 per mille or less.} 
\label{fig:bond_dim}
\end{figure}

\begin{figure}[t!]
    \centering
    \includegraphics[width = 8.6cm]{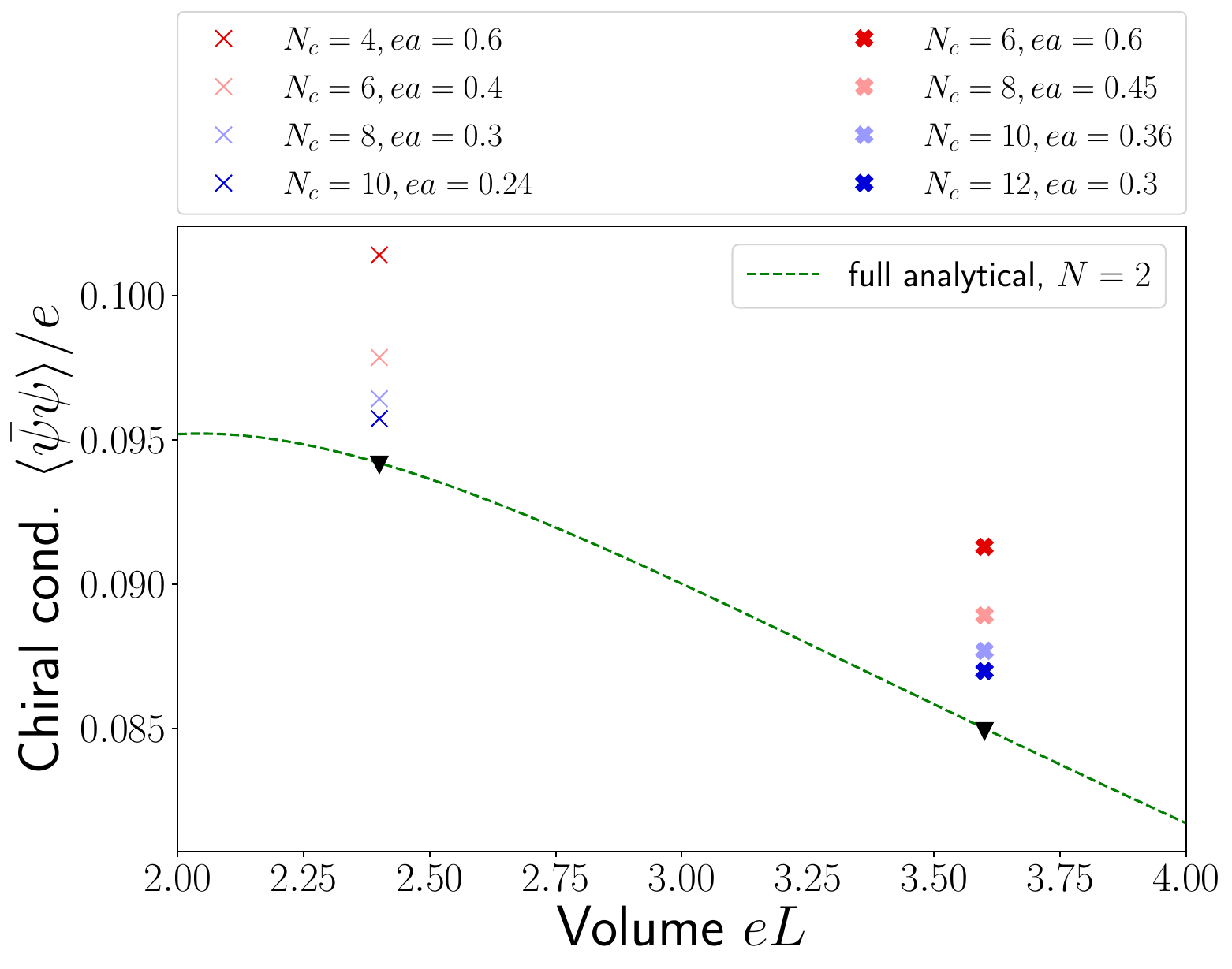}
\caption{
\textbf{Chiral condensate scaling for different lattice spacings at fixed physical volume.} The numerical value of the chiral condensate is plotted for different lattice sizes and lattice spacings so that the physical volume is fixed (procedure done for two different physical volumes, $eL = 2.4$ and $eL = 3.6$). The numerical results converge toward the continuum prediction as the lattice spacing $ea$ decreases. Fitting the finite lattice spacing values as explained in the text allows us to calculate the continuum value of the chiral condensate for both physical volumes (black triangles).} 
\label{fig:lattice_spacing}
\end{figure}

In the main text, the results shown in FIG.~2 are calculated for a fixed bond dimension $\chi$, lattice size $N_s$ and lattice spacing $ea$. In order to extract continuum physics, one ideally needs to extrapolate to infinite bond dimension ($\chi \rightarrow \infty$) and zero lattice spacing $ea\rightarrow 0$, while keeping the physical volume $eL = eaN_s$ constant. The former extrapolation ensures that the result is independent of the MPS approximation and the latter that the result is independent of lattice artifacts. 
In this section, we investigate the effects that a finite bond dimension $\chi$ of the MPS ansatz and finite lattice spacing $ea$ have on the value of the chiral condensate. 

We calculate the chiral condensate at fixed physical volume and lattice spacing for five different values of the maximal bond dimension during the DMRG procedure ($\chi = 200, 250, 300, 350,$ and $400$). In FIG.~\ref{fig:bond_dim}, we show the bond dimension extrapolation results for four different combinations of lattice size and lattice spacing. Following the prescription of Ref.~\cite{Banuls2016}, we estimate the order of magnitude of the error due to finite bond dimensions on the chiral condensate $\Delta\langle\bar{\psi}\psi\rangle$ as the difference in the results for the largest two values, $\chi = 350$ and $400$.

In the remainder of this section, we comment on the extrapolation to the zero lattice spacing limit ($ea\rightarrow 0$ and $N_c\rightarrow \infty$) at fixed finite volume $eL = eaN_s$. 
One way to calculate the continuum value is to fit a polynomial to the results for the chiral condensate obtained at fixed volume and different lattice spacings, 
\begin{align}
    \langle\bar{\psi}\psi\rangle_L(ea) = x_L + y_L(ea)^{z_L},
\label{eq:fitting_fct}
\end{align}
where $x_L,y_L,$ and $z_L$ are fitting parameters. The subscript $L$ indicates that, in general, they are functions of the physical volume $eL$. 

In FIG.~\ref{fig:lattice_spacing}, the value of the chiral condensate is plotted for two different values of the physical volume ($eL = 2.4$ and $3.6$) and for varying lattice spacings. Fitting the polynomial of Eq.~\eqref{eq:fitting_fct} allows us to calculate $x_L, y_L,$ and $z_L$ at each physical volume, and thus to extrapolate to the continuum value of the chiral condensate by evaluating the function $\langle\bar{\psi}\psi\rangle_L(0)$. We obtain the following polynomials at $eL = 2.4$ and $eL = 3.6$:
\begin{align}
\langle\bar{\psi}\psi\rangle_{2.4}(ea) &= 0.0941 + 0.0170\times(ea)^{1.66}\notag\\
\langle\bar{\psi}\psi\rangle_{3.6}(ea) &= 0.0849 + 0.0146\times(ea)^{1.62}.
\end{align} 
Strikingly, the extrapolation yields precise agreement with the analytics (relative error of $10^{-3}$).
In comparison, the relative error for the points in FIG.~2 in the main text, obtained at small but fixed lattice spacing, is of the order of $10^{-2}$, which we deem sufficient for our purposes. Notably, the lattice artifacts can be identified as the dominant source of discrepancy between numerics and analytics.

\section{Fidelity susceptibility}
\label{app:fid_suscept}
In FIG.~1 of the main text, we show the fidelity susceptibility of the ground state as a function of the $\theta$ angle. 
It is defined as
\begin{align}
    \mathcal{F}^{(2)}_0(\theta) = \frac{\partial^2}{\partial\delta^2}|\braket{\psi(\theta)|\psi(\theta+\delta)}|^2\vert_{\delta = 0}.
\end{align}
Taking an information theoretical point of view, the fidelity susceptibility quantifies the change of the distance in the Hilbert space under an infinitesimal change of the control parameter (in our case, the topological angle $\theta$)~\cite{Zanardi2007,Gu2010}. 
The possibility to use this quantity is another advantage of the Hamiltonian lattice formulation assumed in this work. 

In the Lehmann representation, it can be written explicitly as  
\begin{align}
    \mathcal{F}^{(2)}_0(\theta) = \sum_{\beta>0} \frac{|\bra{\psi(\theta)_\beta} H_\theta \ket{\psi(\theta)} |^2}{(E_\beta(\theta)-E_0(\theta))^2}\,, 
\end{align}
where $\ket{\psi(\theta)_\beta}$, $\beta>0$, are the excited energy eigenstates at topological angle $\theta$ with energy $E_\beta(\theta)$, and $E_0(\theta)$ is the ground-state energy. In our case, the driving Hamiltonian is $H_{\theta} = \frac{e^2a}{2\pi}\sum_nS^z_n$.
In spirit, it is therefore similar to the static susceptibility, which measures the change of the ground-state energy as a Hamiltonian parameter is modified, but---as the above formula shows---the fidelity susceptibility is more sensitive toward small gaps in the ground state, as the static susceptibility has only a linear dependence in the denominator on $(E_\beta(\theta)-E_0(\theta))$ \cite{You2015}. 

For this reason, and because it does not require any \textit{a priori} knowledge of order parameters, the fidelity susceptibility has become a valuable tool for identifying quantum phase transitions, characterized by universal divergencies~\cite{Gu2010,Hauke2010,Wang2015} (though, as for other quantifiers, the characterization of Berezinskii--Kosterlitz--Thouless transitions is subtle \cite{Sun2015,You2015}). 
In this work, we exploit these features to distinguish an avoided crossing, where the fidelity susceptibility will show only a broad peak, from a true level crossing. In the latter case, as the properties of the ground state on the left of the crossing point are drastically different from the ones on the right, we expect a deltalike peak, as we indeed observe in our numerics.

In practice, to circumvent issues with numerical derivatives, we compute the discretized version of the second derivative, 
\begin{align}
    \mathcal{F}^{(2)}_0(\theta,\delta) = \frac{1}{\delta^2}\big[&|\braket{\psi(\theta)|\psi(\theta+\delta)}|^2\notag\\+&|\braket{\psi(\theta)|\psi(\theta-\delta)}|^2-2\big],
\end{align}
for a sufficiently small $\delta = 0.025$. Further decrease of $\delta$ shows a narrowing of the peak and exploding of the amplitude in case of flavor-twisted boundary conditions and no significant change in case of flavor-independent boundary conditions. The fact that the fidelity susceptibility behaves differently for fbc with respect to sbc means that the periodicity in $\theta$ in the former case is $4\pi$, whereas in the later case it is $2\pi$.

\end{document}